\documentclass[aps,prl,twocolumn,reprint,superscriptaddress,nofootinbib,preprintnumbers,longbibliography,floatfix]{revtex4-2}

\usepackage[english]{babel}
\usepackage[utf8]{inputenc}
\usepackage{amssymb}
\usepackage{amsmath}
\usepackage{color}
\usepackage{hyperref}
\usepackage{bbm}
\usepackage{epsfig}
\usepackage{multirow}
\usepackage[percent]{overpic}

\usepackage[normalem]{ulem}

\usepackage[dvipsnames]{xcolor}
\usepackage{xspace}
\usepackage{orcidlink}

\newcommand{\e}{\varepsilon}
\newcommand{\s}{\sigma}

\newcommand{\SpinJ}{\mathcal{J}}

\newcommand{\vk}{\mathbf{k}}

\newcommand{\ket}[1]{\left| {#1} \right\rangle }
\newcommand{\bra}[1]{\left\langle {#1} \right| }
\newcommand{\avg}[1]{\left\langle {#1} \right\rangle }

\newcommand{\op}[1]{\hat{#1}}

\newcommand{\ck}{\op{c}^\dagger}

\newcommand{\ca}{\op{c}^{}}

\newcommand{\beq}{ \begin{equation} } 
\newcommand{\eeq}{ \end{equation} }
\newcommand{\beqa}{\begin{eqnarray}}
\newcommand{\eeqa}{\end{eqnarray}}
\newcommand{\nn}{\nonumber}

\newcommand{\ie}{\textit{i.e.}}
\newcommand{\eg}{\textit{e.g.}}

\newcommand{\logicsec}[1]{\color{purple}\emph{#1}\color{black}.---}
\newcommand{\fig}[1]{Fig.~\ref{fig:#1}}

\newcommand{\eq}[1]{Eq.~(\ref{#1})}

\hypersetup{
    bookmarks=false,        
    colorlinks=true,        
    linkcolor=red,        
    citecolor=blue,        
    filecolor=blue,      
    urlcolor=blue
}

\begin{document}
\title{Inverted anisotropy of the partially screened magnetic impurity}

\author{Krzysztof P. W{\'o}jcik\,\orcidlink{0000-0002-8201-1824}}
\email{kpwojcik@ifmpan.poznan.pl}
\affiliation{Institute of Molecular Physics, Polish Academy of Sciences, 
			 Smoluchowskiego 17, 60-179 Pozna{\'n}, Poland}
			 
\author{Micha\l{} P. Kwasigroch\,\orcidlink{0000-0002-6613-2183}}
\email{mpk32@cam.ac.uk}
\affiliation{Trinity College, Trinity Street, Cambridge CB2 1TQ, United Kingdom}
\affiliation{Department of Mathematics, University College London, 
             Gordon St., London WC1H 0AY, United Kingdom}

\date{\today}

\begin{abstract}
We investigate a single magnetic impurity in the presence of strong spin-orbit coupling and 
single-ion anisotropy.
We show that at sufficiently strong coupling there exists a finite temperature window, before the moment is completely screened,
where the magnetic anisotropy of the system flips: the hard-axis becomes the easy-axis or vice versa. We derive this rigorously for a single impurity using numerical renormalization group calculations as well as Nozi\'{e}res' strong-coupling limit and discuss its relevance to heavy-fermion compounds which order magnetically along the hard-direction. We show that the coexistence of Curie-like response and Kondo fluctuations is stabilised along the initially hard direction leading to the anisotropy switch.

\end{abstract}

\maketitle

\logicsec{Introduction}
The scattering of conduction electrons off magnetic impurities and its logarithmic growth with temperature was first characterized by Kondo in his seminal work of 1964 \cite{Kondo}. It set off a chain of discoveries and breakthroughs that continues to this day. Kondo's insight was quickly followed by Anderson and Yuval who showed that the Kondo model flows to the strong coupling fixed point as temperature (or the energy scale) is lowered \cite{Anderson1970Jun}. Wilson's numerical renormalization group (NRG) characterized this flow exactly showing that the impurity is fully screened in the ground state as a singlet state with zero angular momentum is formed \cite{WilsonNRG}. Nozi{\' e}res showed that a Fermi liquid forms at the fixed point as the screened impurity essentially decouples from the rest of the conduction electrons \cite{NozieresBlandin}. Early on, it was also realized by Coqblin and Schrieffer that for rare-Earth ions such as Ce or Yb, where the spin-orbit coupling is strong, the interaction between the local moment and conduction electrons that conserves total angular momentum is more appropriate, generalizing the $\mathrm{SU}(2)$-symmetric Kondo model to the $\mathrm{SU}(N)$-symmetric Coqblin-Schrieffer (CS) model \cite{Coqblin1969Sep}. A plethora of exotic generalizations, differing in the impurity degeneracy, 
the number and nature of screening channels, and forms of anisotropy has been recognized \cite{CoxZawadowski,Otte2008Nov,Golubeva2017Mar,Kogan2021Mar}.

The above advances on the single magnetic impurity were soon followed by the discovery of heavy-fermion (HF) compounds where the scattering of conduction electrons from an entire lattice of impurities becomes coherent at low temperatures giving rise to a Fermi surface with an enhanced volume \cite{Hewson_book}. The participation of the local moments in the Fermi surface mixes local and itinerant degrees of freedom and leads to rich behavior and the associated phenomena are continuing to perplex us even today: unconventional superconductivity, metamagnetism, hard-direction ordering, and many more, see for example Refs. \cite{Aoki_Review,Miyake,Brando}.

The vast majority of the aforementioned single-impurity works was focused on analysis of the 
fixed points of the renormalization group (RG) transformation. Even unstable 
fixed points play a pivotal role in the shaping of universal properties of 
impurity systems, often remaining relevant at intermediate temperature scales. 
Further, the RG flow between the fixed points may exhibit some generic 
features representative of many materials. This can be viewed as the backbone
of the two-fluid description of the Kondo lattice \cite{Nakatsuji2004Jan,Yang2008Mar}, 
which proposes a mixture of residual local moments with a partially formed heavy Fermi 
liquid as a generic picture of a HF compound at intermediate temperature scales. The liquid component grows as temperature is lowered whereas the residual local moments diminish \cite{Nakatsuji2004Jan,Yang2008Mar}. This 2-fluid model has been very successful in characterizing the behavior of heavy-fermion compounds at non-zero temperature, for example by linking the anomalous Knight shift to the logarithmically rising susceptibility of the growing heavy-fermion liquid component \cite{Schmallian_knight}. A particularly interesting scenario is the situation where the system magnetically orders before the residual local moments have vanished entirely. The local moments have partially enlarged the Fermi surface and are entangled with electronic degrees of freedom. In RG language, they are no longer the bare local moments of the original theory but renormalized ones with renormalized effective interactions. The magnetically ordered phase is likely to have very different properties to if bare impurities had ordered.

One such property is the the preferred direction of order and how it compares with the preferred direction of the bare moments. In fact, in a number of lanthanide and actinide heavy-fermion compounds a switch in the magnetic anisotropy of the system  has been observed as temperature is lowered: the hard-axis becomes the easy-axis and vice versa, which leads to magnetic order (if present) that is perpendicular to the easy direction at high temperature \cite{Brando}. Several works have highlighted the strong correlation with the Kondo effect \cite{Brando, Kwasigroch2022Jun,Vojta&Brando}, and recently it had been found that in a few  heavy-fermion families the temperature at which the anisotropy of the uniform susceptibility flips is directly proportional to the temperature at which Kondo scattering becomes coherent \cite{Scott2025Dec}.

Several explanations for the easy-axis switch have been put forward: 
phenomenological~\cite{Vojta&Brando,Jin2025}, 
perturbative~\cite{Kwasigroch2022Jun}, 
or mean-field~\cite{Scott2025Dec}. 
The mechanism proposed in Ref.~\cite{Scott2025Dec} relies on local correlations between conduction electrons and impurities, in the sense that it can cause the latter to order either ferromagnetically or antiferromagnetically along the hard-axis. For this reason, perhaps the mechanism can also be seen at the level of a single impurity. This serves as the motivation for the present Letter, where we go beyond the mean-field or perturbative treatment of previous works and rigorously show using NRG calculations, as well as Nozi\`{e}res strong-coupling limit, that the easy and hard axes switch in the partially screened state for strong enough Kondo coupling. The partially screened state here is the direct analogue of the 2-fluid state in the lattice. There is still a non-zero local moment, and non-vanishing Curie response, but also strong Kondo fluctuations that cause the impurity to hybridize and share the moment with the electron fluid.

\logicsec{Model}%
In the presence of strong spin-orbit coupling, a magnetic impurity is described by the following Coqblin-Schrieffer model \cite{Coqblin1969Sep}  with additional anisotropy term,
\beqa
H_{\rm CS} 	&=& \sum_{k m m'} \e_{k mm'} \ck_{k m} \ca_{k m'}
	- \sum_a B_a \op\SpinJ_a 	+ D \op\SpinJ_z^2 
	\nn\\&&\
	+ \frac{J}{N_{\rm BZ}}  \sum_{kk' mm'}\!\! \left( \! \op{X}_{m'm} \!\! - \!\frac{\delta_{mm'}}{2j+1} \!\right) \ck_{k m} \ca_{k' m'} .
\label{H_CS}
\eeqa
Here, $\ca_{\vk m}$ annihilates conduction band electron with momentum $k$, 
total angular momentum $j$ and its projection on anisotropy axis $m$.
Hubbard operators $\op{X}_{m'm} \equiv \ket{m'}\bra{m}$ act only on impurity degrees of freedom, 
changing the angular momentum component.
$\SpinJ_a$ denotes a $4\times4$ matrix of $a$-th component of total angular momentum,
with $a$ running through real-space axes $x$, $y$, $z$, while $\op\SpinJ_a$
is an operator of a corresponding projection of impurity total angular momentum. 
We keep the exchange coupling $J$ isotropic, implementing anisotropy 
at the level of bare impurity energy through anisotropy axial constant $D$. 
$N_{\rm BZ}$ denotes the number of $\vk$-points in the Brillouin zone and 
normalizes the $\vk$ summations in the exchange term.
We also follow a convention that keeps the exchange interaction traceless \cite{Coqblin1969Sep},
to disentangle the effects of potential scattering.
The band dispersion term in generic magnetic field $\vec{B}$ has a form 
$\e_{\vk mm'} = \e_{\vk} + B_a (\SpinJ_a)_{mm'}$. We have incorporated 
the gyromagnetic factor into $\vec{B}$ through the choice of units.
Without loss of generality we parametrize $B_x = B\sin(\alpha)$, $B_z = B\cos(\alpha)$, 
and take $B_y=0$. 
We assume rectangular conduction band of half-width $W$. 
Although, we will focus on the a magnetic impurity with total angular momentum $j=3/2$, we have observed qualitatively similar results in the strong-coupling limit for other values of $j$.

We are mostly interested in anisotropy of impurity magnetic susceptibility, defined as 
\mbox{$\chi_a = \lim_{B_a\to 0} \avg{\op\SpinJ_a}/B_a$}, when the magnetic field is uniformly
applied to the whole system, cf.~\eq{H_CS}. We calculate that directly, applying a small magnetic 
field $B=10^{-5}W$. 
It is sufficient to study impurity response (not the total one), as it is qualitatively similar, 
as long as the field acts uniformly everywhere. 
We identify the easy-axis switch and the inversion of the effective anisotropy with a crossing of the magnetic susceptibilities for fields along two perpendicular directions $\Delta \chi=\chi_z - \chi_x$.
We also calculate impurity contribution to entropy \cite{NRG_RMP}, $\Delta S^{\rm imp}$, 
\ie{} the difference in entropy between the system with and without impurity.
This calculation is done in the absence of magnetic field, $B=0$.

\begin{figure}[bt!]
\includegraphics[width=\linewidth]{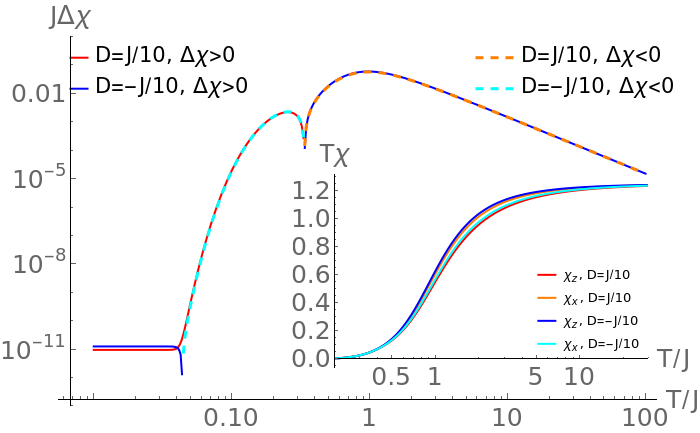}
\caption{The anisotropy of the magnetic susceptibility of the impurity, 
        $\Delta\chi$ in the strong coupling limit $J/W \gg 1$,
        calculated using finite $B=10^{-4}J$. 
        Note the logarithmic scale on both axes.
        Inset: actual $\chi_z$ and $\chi_x$ as functions of $T$.
        }
\label{fig:SClimit}
\end{figure}

\logicsec{Strong-coupling limit}%
We begin with the Nozeries strong-coupling limit $J\gg W$, which provides a simple and intuitive picture of the partially screened impurity. Then, the kinetic energy term can be neglected.
Defining $\op\psi_m = N_{\rm BZ}^{-1/2} \sum_{\vk} \ca_{\vk m}$ one then
reduces the Hamiltonian, \eq{H_CS}, to a $2$-site problem that could be solved 
exactly. 
The results for $\chi_z$ and $\chi_x$ are presented in the inset of \fig{SClimit},
together with the susceptibility anisotropy $\Delta\chi$ in the main plot.
At elevated temperatures, $T \gg J, |D|$, the susceptibilities approach a free-impurity 
value, $\chi_0 = j(j+1)/(3T) = 5/(4T)$. Susceptibility anisotropy grows when the temperature
is decreased, and its sign remains intuitively understandable.
Below $T\approx 0.3J$, the situation changes. For $D>0$, the hard-axis becomes the easy-axis, as the parallel susceptibility exceeds the perpendicular one.
On the contrary, for $D<0$, the easy-axis becomes the hard-axis, as the parallel susceptibility becomes smaller than the perpendicular one. In this case, below $T \approx 0.05J$ a sign change of $\Delta\chi$ occurs again, leading to the restoration of the original high-temperature aniostropy.

We can understand the reversal of the anisotropy as follows. 
For $D=0$, the fully screened \emph{strong coupling} ground state is given by
\begin{align}
\ket{\rm SC} =\frac{1}{2}\sum\limits_{\substack{\s \in S_4 \\ \s(4)>\s(3)>\s(2)}} 
{\rm sgn} (\s)
\ck_{\s(2)}\ck_{\s(3)}\ck_{\s(4)} | \s(1)\rangle,
\end{align}
where we sum over permutations $\s(1)\ldots\s(4)$ of the angular momentum projections 
$\left\{-\frac{3}{2},-\frac{1}{2},\frac{1}{2},\frac{3}{2} \right\}$,
and $\ket{m}$ denotes a state with empty conduction band site and impurity with 
angular momentum projection $m$. 
The four-fold degenerate (first-excited) partially screened quartet
components are given by
\begin{align}
| E(m)\rangle =\frac{1}{\sqrt{3}}\sum\limits_{\substack{\s \in S_3^m \\ \s(3)>\s(2)}} 
{\rm sgn} (\s)
\ck_{\s(2)}\ck_{\s(3)}| \s(1)\rangle,
\end{align}
where $S_3^m$ is a set of permutations of the three of 
$\left\{-\frac{3}{2},-\frac{1}{2},\frac{1}{2},\frac{3}{2}\right\}$ different than $m$. 
For instance, the partially screened state with the maximum polarization along the $z$-axis is
\begin{align}
\left|E\left(\frac{3}{2}\right) \right\rangle 
= 
\frac{1}{\sqrt{3}}
\left(\ck_{1/2}\ck_{3/2}\ket{-\frac{1}{2}}-
\ck_{-1/2}\ck_{3/2} \ket{\frac{1}{2}} +\right.
\nonumber
\\
\left.
\ck_{-1/2}\ck_{1/2} \ket{\frac{3}{2}}\right)
\end{align}
We can see that Kondo fluctuations reduce the expected value of $m$ 
on the impurity as the moment is shared by the neighboring electrons. 
They also reduce the expected value of $m^2$, as the impurity fluctuates 
between all possible $m$-states except for $m=-\frac{3}{2}$. 
Thus, fluctuations cause the polarized impurity moment to bend towards 
the \mbox{$xy$-plane}. 
Note that the expected value of $m$ for conduction band electrons 
(per electron) is the same same as for the impurity, $\avg{m_c}/2 = \avg{m_f} = 1/2$,
and yet they are on average antialligned, $\avg{m_c m_f} < 0$. This entanglement is the core of the mechanism and of the analogy to the $2$-fluid model, where the impurities are locally antialligning with the conduction electrons but also polarising the same way in a field as part of the heavy-fermion liquid. 
We thus conclude that if an external field breaks the degeneracy of the local-moment subspace (or in the case of spontaneous symmetry breaking on a lattice), Kondo fluctuations lower the expected $m$ of the impurity but importantly also $m^2$ below the isotropic value of  $j(j+1)/3$.  The addition of weak single-ion anisotropy stabilises the coexistence of broken local-moment degeneracy and Kondo fluctuations along the hard-axis, i.e., the energy difference between the state aligned along the hard and easy axes is
\begin{align}
     D \left\langle\op\SpinJ_z^2 -\op\SpinJ_x^2 \right\rangle = 
     \frac{3}{2}\left( \langle m^2\rangle-\frac{j(j+1)}{3}
     \right)<0,
\end{align}
where we have taken $D>0$ so that $z$ is the hard axis.

The most important lesson one draws from the strong-coupling limit is that strong exchange 
coupling may lead to counter-intuitive behavior of the susceptibility anisotropy at intermediate
temperatures, $T\sim J$. While for $T\to 0$ and $T\to\infty$ the impurity response
becomes isotropic, in the region of its maximal absolute value its sign does not have to 
follow from the sign of anisotropy constant $D$. 
Instead, in the crossover regime, where the singlet is only partially formed, the susceptibility 
along the high-temperature hard direction is in fact larger.

\begin{figure}[tb!]
\begin{overpic}[width=0.8\linewidth]{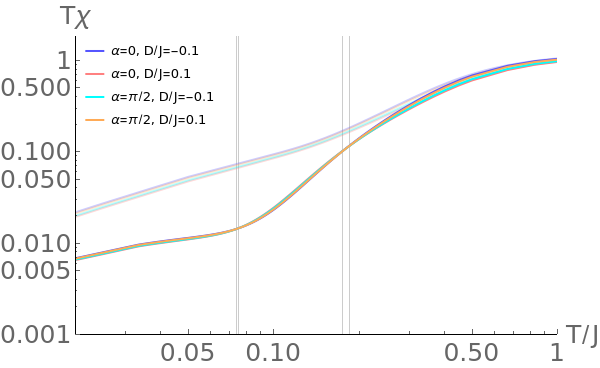}
 \put(1,55){(a)}
 \put(60,8){\includegraphics[width=0.4\linewidth]{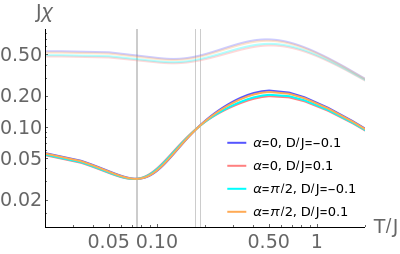}}
\end{overpic}\hspace{0.2\linewidth}\\
\begin{overpic}[width=\linewidth]{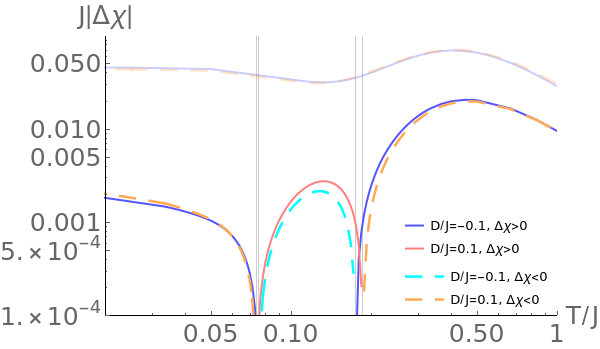}
 \put(1,55){(b)}
\end{overpic}
\caption{(a) The susceptibility of the impurity (times temperature) in a uniform field 
         in the case of the CS model with $j=3/2$ and $\rho J=3$ (light lines indicate $\rho J =1$ case), 
         calculated by NRG.          
         Inset: actual $\chi$ as function of $T$.
         (b) Corresponding difference $\Delta\chi$.
         }
\label{fig:chi}
\end{figure}

\logicsec{NRG results}%
Motivated by the strong coupling results we proceed to analyze a more realistic 
situation of a finite exchange-to-bandwidth ratio, when the kinetic energy term
affects the situation, in particular leading to the Kondo effect. 
We perform NRG calculations using discretization parameter $\Lambda=4$ 
and keeping $N_{\rm kept}\approx 1000$ states at each NRG step \cite{NRG_RMP,fnrg}.
The resulting susceptibilities, obtained from results with a finite magnetic field 
$B=10^{-5}W$, are presented in \fig{chi} for two values of the exchange coupling: 
$\rho J = 1$ and $\rho J =3$. 

First of all, we see that such strong couplings lead to very quick suppression 
of the local impurity moment, $T\chi$. It happens already at $T\sim J$, \ie{} 
at bare exchange energy scale. For $\rho J=1$ we see all susceptibilities 
possess Curie characteristics at $T \gg J$ and van Vleck ones at $T \ll J$. 
However, the most intriguing is the crossover regime, where they all exhibit
a minimum. In the case of $\rho J=3$ it leads to a Curie-like regime at $T\lesssim J$. 
Note that such minimum is not present for anisotropic model in the scaling regime, 
$J\ll W$ \cite{Rajan1983Jul}.
In this regime, the strong-coupling solution is realized and $\Delta\chi$
indeed changes sign. This only happens for strong enough exchange couplings,
$\rho J \gtrsim 1.5$, and is only weakly dependent on $D$.

\begin{figure}[tb!]
\includegraphics[width=\linewidth]{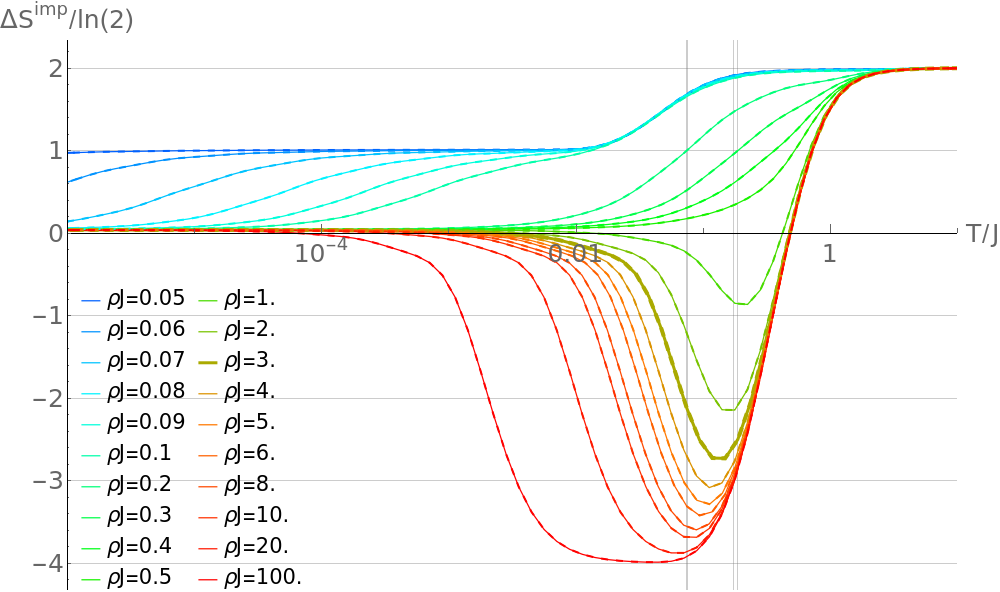}
\caption{Impurity contribution to entropy, $\Delta S^{\rm imp}$, as a function of $T$
        for indicated values of $J$ and $D=J/10$ (solid lines) and $D=-J/10$ (dashed lines).
        Vertical lines indicate positions of sign-changes of $\Delta\chi$ for $\rho J=3$
        (thick line).
        }
\label{fig:entropy}
\end{figure}

It should be stressed that the regime of the partially screened moment (reduced Curie moment $T\chi$) and inverted anisotropy
(inverted sign of $\Delta\chi$) does not lie at the Kondo 
fixed point which is reached as $T\rightarrow 0$. This is most evidently seen by studying impurity contribution to
entropy, $\Delta S^{\rm imp}$, cf.~\fig{entropy}.  
The entropy contribution is calculated from the partition function of the whole 
Wilson chain (a moving average over a temperature corresponding on a log-scale
to a factor $3.16$ was used to reduce discretization artifacts).
In Kondo systems $\Delta S^{\rm imp}$ tends to be a monotonic function of $T$, 
decreasing from the logarithm of the number of impurity degrees of freedom, $\ln(N)$,
at large $T$, to $0$ at $T\to 0$. This, however, only happens for $\rho J \ll 1$,
when $\Delta\chi(T)$ has no sign changes. For strong enough $J$ a regime of negative
$\Delta S^{\rm imp}$ develops, indicating local singlet formation instead of 
usual Kondo cloud formation,
and suggesting a possible instability for a sufficient density of impurities \cite{KWJKPW}.
The maximal reduction of entropy by the impurity, $\Delta S^{\rm imp} = -4\ln(2)$,
corresponds to effectively removing one $k$-point from the Brillouin zone for 
a CS model with $N=4$ angular momentum projections.

We briefly note here that for $\rho J \leq 0.1$, the impurity contribution to entropy
is monotonically decreasing and exhibits a two-stage decay. 
First (at $T\sim |D|=J/10$) $\Delta S^{\rm imp}$ is reduced when the anisotropy 
term renders a fraction of impurity degrees of freedom thermally inaccessible, 
and later (at $T=T_K \sim W e^{-1/2\rho J}$) it is suppressed to $0$ due to full 
Kondo screening, characteristic of the CS model.
For intermediate $\rho J$, $0.2\leq \rho J \leq 0.5$, not only does $\Delta S^{\rm imp}$ remain monotonically decreasing, but it also exhibits a single decay at $T\sim T_K > |D|$, leaving no residual moments. 
It is only for $\rho J \gtrsim 1$ when the negative $\Delta S^{\rm imp}$ occurs
and the anisotropy changes sign.

\logicsec{Discussion}%
So far, the proposed explanations of the inverted anisotropy and order along the direction that is hard at high temperature include fluctuation-based mechanisms \cite{Brando}, or an RKKY exchange that happens to be stronger along the high-temperature hard axis \cite{Vojta&Brando, Jin2025,Scott2025Dec}. The anisotropy in the RKKY could be caused by the relative placement of the magnetic ions \cite{Jin2025}, which would be very lattice specific, or anisotropic Kondo exchange \cite{Kwasigroch2022Jun,Scott2025Dec}. The main conundrum is why the exchange should be stronger along the hard axis. Perturbative RG calculations for a single impurity show that local single-ion anisotropy generates an effective Kondo exchange (for the renormalized moments) that is stronger along the hard axis \cite{Kwasigroch2022Jun}. This would then give stronger RKKY exchange between the impurities on a lattice. Indeed, mean-field calculations for an underscreend spin-1 Kondo lattice reaffirm this picture by showing that the generated exchange between the residual moments is stronger along the high-temperature hard direction \cite{Scott2025Dec}. The key reason for the stronger exchange along the hard direction and consequently lower energy of a state that orders along this direction is that single-ion anisotropy stabilizes the coexistence of magnetized residual local moments and Kondo hybridization along the hard direction, i.e., the polarized 2-fluid state. In the presence of symmetry breaking that polarizes the conduction electrons, the residual local impurity moments try to antialign with the conduction electrons (see $\langle m_cm_f\rangle <0 $ in the previous discussion). However, through Kondo hybridization, the impurities also become part of the itinerant Fermi liquid, and if it polarizes, they polarize in the same direction as the conduction electrons (see $\langle m_f \rangle ,\langle m_c \rangle >0$ in the previous discussion). This inherent frustration present in the 2-fluid state, when the impurities are part of both the heavy-fermion fluid and the local moment one, is relieved along the hard direction in the presence of single-ion anisotropy. 

Unlike the previous perturbative and mean-field approaches, the numerical results presented in this Letter are based on a direct calculation 
with the aid of the most reliable non-perturbative single-impurity solver, 
NRG. Therefore, one can confidently conclude that for sufficiently strong exchange
coupling, the susceptibility anisotropy of an impurity to a uniform magnetic 
field changes sign in a temperature window below $T\sim J$. 
The study of impurity contribution to entropy clearly shows that this regime
does not correspond to incomplete conventional Kondo cloud formation, but rather
resembles static local singlet formation (which is not the same and can be 
identified from $\Delta S^{\rm imp}$). 
The relevance of these results for the phenomenology motivating our study
is yet to be fully established.

Firstly, one might worry that the effect seems to require $\rho J > 1$,
which for a rectangular band corresponds to an exchange coupling of the order
of the conduction bandwidth. However, in realistic scenarios the conduction 
band density of states possesses many peaks, \eg{} caused by Van Hove 
singularities, such that the role of the bandwidth might be played by a 
characteristic width of a relevant spectral feature. Thus we believe 
this is not a major concern.

More significant is the question how far a single-impurity calculation can 
be trusted in capturing the essence of correlated lattice physics. 
Since a generic correlated lattice problem cannot be solved without 
uncontrolled approximations, this point requires some attention. 
Clearly, the phenomenon does not involve the partial formation of the conventional Kondo cloud which takes place as the impurity contribution to entropy tends to zero at the usual Kondo fixed point.
However, it does correspond to the partial screening of the impurity moment: partially reduced Curie moment $T\chi$, and on the mean-field level, partial hybridization of the impurity with the conduction electrons. We believe this is a direct analogue of the 2-fluid state seen at intermediate temperatures in heavy-fermion compounds. Our results show that the partially screened and magnetized impurity experiences an inverted anisotropy -- it has a lower energy (for the same applied field) along the high-temperature hard axis. Similarly, on a lattice it has been found that a 2-fluid state has a lower energy when polarized along the high-temperature hard direction.

The present results for the single-impurity could be most directly connected to the lattice problem, at least on a mean-field level, via a dynamical mean-field theory \cite{DMFT_RMP}. We expect that the enhanced magnetic response of a single impurity along the high-temperature hard direction would give magnetic order along this axis, at least for sufficiently strong Kondo coupling and sufficiently spaced impurities. We postpone the detailed analysis to future work.

\logicsec{Summary}%
We have analyzed a single-impurity Coqblin-Schrieffer model with single-ion 
anisotropy and rigorously concluded that for strong enough exchange coupling and 
at temperatures of the order of that coupling the system exhibits an intriguing regime of partially
formed local singlet, where the anisotropy of the magnetic susceptibility is inverted
with respect to the bare model, yet still retains some Curie character.
We draw an analogy with the behavior of many heavy-fermion compounds,
which at low temperatures order along the magnetically hard direction.
We therefore propose the physical mechanism of both to be the same, based on
exchange (Kondo) fluctuations. 
We believe that our detailed NRG analysis might help to find more materials sharing these characteristics.
We hope to stimulate further efforts towards the understanding of hard-direction 
ordering and its utilization in quantum material design.

\begin{acknowledgments}
We greatfully acknowledge stimulating discussions with Ewan Scott.
KPW was supported by National Science Centre in Poland 
through grant no.~2023/51/D/ST3/00532
and acknowledges computing time at Pozna\'n Supercomputing 
and Networking Center.
All data used in the publication has been made available 
at the Zenodo repository...
\end{acknowledgments}


\begin{thebibliography}{25}%
\makeatletter
\providecommand \@ifxundefined [1]{%
 \@ifx{#1\undefined}
}%
\providecommand \@ifnum [1]{%
 \ifnum #1\expandafter \@firstoftwo
 \else \expandafter \@secondoftwo
 \fi
}%
\providecommand \@ifx [1]{%
 \ifx #1\expandafter \@firstoftwo
 \else \expandafter \@secondoftwo
 \fi
}%
\providecommand \natexlab [1]{#1}%
\providecommand \enquote  [1]{``#1''}%
\providecommand \bibnamefont  [1]{#1}%
\providecommand \bibfnamefont [1]{#1}%
\providecommand \citenamefont [1]{#1}%
\providecommand \href@noop [0]{\@secondoftwo}%
\providecommand \href [0]{\begingroup \@sanitize@url \@href}%
\providecommand \@href[1]{\@@startlink{#1}\@@href}%
\providecommand \@@href[1]{\endgroup#1\@@endlink}%
\providecommand \@sanitize@url [0]{\catcode `\\12\catcode `\$12\catcode
  `\&12\catcode `\#12\catcode `\^12\catcode `\_12\catcode `\%12\relax}%
\providecommand \@@startlink[1]{}%
\providecommand \@@endlink[0]{}%
\providecommand \url  [0]{\begingroup\@sanitize@url \@url }%
\providecommand \@url [1]{\endgroup\@href {#1}{\urlprefix }}%
\providecommand \urlprefix  [0]{URL }%
\providecommand \Eprint [0]{\href }%
\providecommand \doibase [0]{https://doi.org/}%
\providecommand \selectlanguage [0]{\@gobble}%
\providecommand \bibinfo  [0]{\@secondoftwo}%
\providecommand \bibfield  [0]{\@secondoftwo}%
\providecommand \translation [1]{[#1]}%
\providecommand \BibitemOpen [0]{}%
\providecommand \bibitemStop [0]{}%
\providecommand \bibitemNoStop [0]{.\EOS\space}%
\providecommand \EOS [0]{\spacefactor3000\relax}%
\providecommand \BibitemShut  [1]{\csname bibitem#1\endcsname}%
\let\auto@bib@innerbib\@empty
\bibitem [{\citenamefont {Kondo}(1964)}]{Kondo}%
  \BibitemOpen
  \bibfield  {author} {\bibinfo {author} {\bibfnamefont {J.}~\bibnamefont
  {Kondo}},\ }\bibfield  {title} {\bibinfo {title} {{Resistance Minimum in
  Dilute Magnetic Alloys}},\ }\href {https://doi.org/10.1143/PTP.32.37}
  {\bibfield  {journal} {\bibinfo  {journal} {Prog. Theor. Phys.}\ }\textbf
  {\bibinfo {volume} {32}},\ \bibinfo {pages} {37} (\bibinfo {year}
  {1964})}\BibitemShut {NoStop}%
\bibitem [{\citenamefont {Anderson}\ \emph {et~al.}(1970)\citenamefont
  {Anderson}, \citenamefont {Yuval},\ and\ \citenamefont
  {Hamann}}]{Anderson1970Jun}%
  \BibitemOpen
  \bibfield  {author} {\bibinfo {author} {\bibfnamefont {P.~W.}\ \bibnamefont
  {Anderson}}, \bibinfo {author} {\bibfnamefont {G.}~\bibnamefont {Yuval}},\
  and\ \bibinfo {author} {\bibfnamefont {D.~R.}\ \bibnamefont {Hamann}},\
  }\bibfield  {title} {\bibinfo {title} {{Exact Results in the Kondo Problem.
  II. Scaling Theory, Qualitatively Correct Solution, and Some New Results on
  One-Dimensional Classical Statistical Models}},\ }\href
  {https://doi.org/10.1103/PhysRevB.1.4464} {\bibfield  {journal} {\bibinfo
  {journal} {Phys. Rev. B}\ }\textbf {\bibinfo {volume} {1}},\ \bibinfo {pages}
  {4464} (\bibinfo {year} {1970})}\BibitemShut {NoStop}%
\bibitem [{\citenamefont {Wilson}(1975)}]{WilsonNRG}%
  \BibitemOpen
  \bibfield  {author} {\bibinfo {author} {\bibfnamefont {K.~G.}\ \bibnamefont
  {Wilson}},\ }\bibfield  {title} {\bibinfo {title} {{The renormalization
  group: Critical phenomena and the Kondo problem}},\ }\href
  {https://doi.org/10.1103/RevModPhys.47.773} {\bibfield  {journal} {\bibinfo
  {journal} {Rev. Mod. Phys.}\ }\textbf {\bibinfo {volume} {47}},\ \bibinfo
  {pages} {773} (\bibinfo {year} {1975})}\BibitemShut {NoStop}%
\bibitem [{\citenamefont {Nozi{\ifmmode\grave{e}\else\`{e}\fi}res}\ and\
  \citenamefont {Blandin}(1980)}]{NozieresBlandin}%
  \BibitemOpen
  \bibfield  {author} {\bibinfo {author} {\bibfnamefont {{\relax
  Ph}.}~\bibnamefont {Nozi{\ifmmode\grave{e}\else\`{e}\fi}res}}\ and\ \bibinfo
  {author} {\bibfnamefont {A.}~\bibnamefont {Blandin}},\ }\bibfield  {title}
  {\bibinfo {title} {{Kondo effect in real metals}},\ }\href
  {https://doi.org/10.1051/jphys:01980004103019300} {\bibfield  {journal}
  {\bibinfo  {journal} {J. Phys.}\ }\textbf {\bibinfo {volume} {41}},\ \bibinfo
  {pages} {193} (\bibinfo {year} {1980})}\BibitemShut {NoStop}%
\bibitem [{\citenamefont {Coqblin}\ and\ \citenamefont
  {Schrieffer}(1969)}]{Coqblin1969Sep}%
  \BibitemOpen
  \bibfield  {author} {\bibinfo {author} {\bibfnamefont {B.}~\bibnamefont
  {Coqblin}}\ and\ \bibinfo {author} {\bibfnamefont {J.~R.}\ \bibnamefont
  {Schrieffer}},\ }\bibfield  {title} {\bibinfo {title} {{Exchange Interaction
  in Alloys with Cerium Impurities}},\ }\href
  {https://doi.org/10.1103/PhysRev.185.847} {\bibfield  {journal} {\bibinfo
  {journal} {Phys. Rev.}\ }\textbf {\bibinfo {volume} {185}},\ \bibinfo {pages}
  {847} (\bibinfo {year} {1969})}\BibitemShut {NoStop}%
\bibitem [{\citenamefont {Cox}\ and\ \citenamefont
  {Zawadowski}(1998)}]{CoxZawadowski}%
  \BibitemOpen
  \bibfield  {author} {\bibinfo {author} {\bibfnamefont {D.~L.}\ \bibnamefont
  {Cox}}\ and\ \bibinfo {author} {\bibfnamefont {A.}~\bibnamefont
  {Zawadowski}},\ }\bibfield  {title} {\bibinfo {title} {{Exotic Kondo effects
  in metals: Magnetic ions in a crystalline electric field and tunnelling
  centres}},\ }\href {https://doi.org/10.1080/000187398243500} {\bibfield
  {journal} {\bibinfo  {journal} {Adv. Phys.}\ }\textbf {\bibinfo {volume}
  {47}},\ \bibinfo {pages} {599} (\bibinfo {year} {1998})}\BibitemShut
  {NoStop}%
\bibitem [{\citenamefont {Otte}\ \emph {et~al.}(2008)\citenamefont {Otte},
  \citenamefont {Ternes}, \citenamefont {von Bergmann}, \citenamefont {Loth},
  \citenamefont {Brune}, \citenamefont {Lutz}, \citenamefont {Hirjibehedin},\
  and\ \citenamefont {Heinrich}}]{Otte2008Nov}%
  \BibitemOpen
  \bibfield  {author} {\bibinfo {author} {\bibfnamefont {A.~F.}\ \bibnamefont
  {Otte}}, \bibinfo {author} {\bibfnamefont {M.}~\bibnamefont {Ternes}},
  \bibinfo {author} {\bibfnamefont {K.}~\bibnamefont {von Bergmann}}, \bibinfo
  {author} {\bibfnamefont {S.}~\bibnamefont {Loth}}, \bibinfo {author}
  {\bibfnamefont {H.}~\bibnamefont {Brune}}, \bibinfo {author} {\bibfnamefont
  {C.~P.}\ \bibnamefont {Lutz}}, \bibinfo {author} {\bibfnamefont {C.~F.}\
  \bibnamefont {Hirjibehedin}},\ and\ \bibinfo {author} {\bibfnamefont {A.~J.}\
  \bibnamefont {Heinrich}},\ }\bibfield  {title} {\bibinfo {title} {{The role
  of magnetic anisotropy in the Kondo effect}},\ }\href
  {https://doi.org/10.1038/nphys1072} {\bibfield  {journal} {\bibinfo
  {journal} {Nat. Phys.}\ }\textbf {\bibinfo {volume} {4}},\ \bibinfo {pages}
  {847} (\bibinfo {year} {2008})}\BibitemShut {NoStop}%
\bibitem [{\citenamefont {Golubeva}\ \emph {et~al.}(2017)\citenamefont
  {Golubeva}, \citenamefont {Sotnikov}, \citenamefont {Cichy}, \citenamefont
  {Kune{\ifmmode\check{s}\else\v{s}\fi}},\ and\ \citenamefont
  {Hofstetter}}]{Golubeva2017Mar}%
  \BibitemOpen
  \bibfield  {author} {\bibinfo {author} {\bibfnamefont {A.}~\bibnamefont
  {Golubeva}}, \bibinfo {author} {\bibfnamefont {A.}~\bibnamefont {Sotnikov}},
  \bibinfo {author} {\bibfnamefont {A.}~\bibnamefont {Cichy}}, \bibinfo
  {author} {\bibfnamefont {J.}~\bibnamefont
  {Kune{\ifmmode\check{s}\else\v{s}\fi}}},\ and\ \bibinfo {author}
  {\bibfnamefont {W.}~\bibnamefont {Hofstetter}},\ }\bibfield  {title}
  {\bibinfo {title} {{Breaking of SU(4) symmetry and interplay between strongly
  correlated phases in the Hubbard model}},\ }\href
  {https://doi.org/10.1103/PhysRevB.95.125108} {\bibfield  {journal} {\bibinfo
  {journal} {Phys. Rev. B}\ }\textbf {\bibinfo {volume} {95}},\ \bibinfo
  {pages} {125108} (\bibinfo {year} {2017})}\BibitemShut {NoStop}%
\bibitem [{\citenamefont {Kogan}\ and\ \citenamefont
  {Shi}(2021)}]{Kogan2021Mar}%
  \BibitemOpen
  \bibfield  {author} {\bibinfo {author} {\bibfnamefont {E.}~\bibnamefont
  {Kogan}}\ and\ \bibinfo {author} {\bibfnamefont {Z.}~\bibnamefont {Shi}},\
  }\bibfield  {title} {\bibinfo {title} {{Poor man{'}s scaling: XYZ
  Coqblin{\textendash}Schrieffer model revisited}},\ }\href
  {https://doi.org/10.1088/1742-5468/abe409} {\bibfield  {journal} {\bibinfo
  {journal} {J. Stat. Mech.: Theory Exp.}\ }\textbf {\bibinfo {volume}
  {2021}}\bibinfo  {number} { (3)},\ \bibinfo {pages} {033101}}\BibitemShut
  {NoStop}%
\bibitem [{\citenamefont {Hewson}(1997)}]{Hewson_book}%
  \BibitemOpen
\bibfield  {number} {  }\bibfield  {author} {\bibinfo {author} {\bibfnamefont
  {A.~C.}\ \bibnamefont {Hewson}},\ }\href
  {https://doi.org/10.1017/CBO9780511470752} {\emph {\bibinfo {title} {{The
  Kondo problem to heavy fermions}}}}\ (\bibinfo  {publisher} {Cambridge
  University Press},\ \bibinfo {address} {Cambridge},\ \bibinfo {year}
  {1997})\BibitemShut {NoStop}%
\bibitem [{\citenamefont {Aoki}\ \emph {et~al.}(2019)\citenamefont {Aoki},
  \citenamefont {Ishida},\ and\ \citenamefont {Flouquet}}]{Aoki_Review}%
  \BibitemOpen
  \bibfield  {author} {\bibinfo {author} {\bibfnamefont {D.}~\bibnamefont
  {Aoki}}, \bibinfo {author} {\bibfnamefont {K.}~\bibnamefont {Ishida}},\ and\
  \bibinfo {author} {\bibfnamefont {J.}~\bibnamefont {Flouquet}},\ }\bibfield
  {title} {\bibinfo {title} {Review of u-based ferromagnetic superconductors:
  Comparison between uge2, urhge, and ucoge},\ }\href
  {https://doi.org/10.7566/JPSJ.88.022001} {\bibfield  {journal} {\bibinfo
  {journal} {Journal of the Physical Society of Japan}\ }\textbf {\bibinfo
  {volume} {88}},\ \bibinfo {pages} {022001} (\bibinfo {year}
  {2019})}\BibitemShut {NoStop}%
\bibitem [{\citenamefont {Miyake}\ \emph {et~al.}(2019)\citenamefont {Miyake},
  \citenamefont {Shimizu}, \citenamefont {Sato}, \citenamefont {Li},
  \citenamefont {Nakamura}, \citenamefont {Homma}, \citenamefont {Honda},
  \citenamefont {Flouquet}, \citenamefont {Tokunaga},\ and\ \citenamefont
  {Aoki}}]{Miyake}%
  \BibitemOpen
  \bibfield  {author} {\bibinfo {author} {\bibfnamefont {A.}~\bibnamefont
  {Miyake}}, \bibinfo {author} {\bibfnamefont {Y.}~\bibnamefont {Shimizu}},
  \bibinfo {author} {\bibfnamefont {Y.~J.}\ \bibnamefont {Sato}}, \bibinfo
  {author} {\bibfnamefont {D.}~\bibnamefont {Li}}, \bibinfo {author}
  {\bibfnamefont {A.}~\bibnamefont {Nakamura}}, \bibinfo {author}
  {\bibfnamefont {Y.}~\bibnamefont {Homma}}, \bibinfo {author} {\bibfnamefont
  {F.}~\bibnamefont {Honda}}, \bibinfo {author} {\bibfnamefont
  {J.}~\bibnamefont {Flouquet}}, \bibinfo {author} {\bibfnamefont
  {M.}~\bibnamefont {Tokunaga}},\ and\ \bibinfo {author} {\bibfnamefont
  {D.}~\bibnamefont {Aoki}},\ }\bibfield  {title} {\bibinfo {title}
  {Metamagnetic transition in heavy fermion superconductor ute2},\ }\href
  {https://doi.org/10.7566/JPSJ.88.063706} {\bibfield  {journal} {\bibinfo
  {journal} {Journal of the Physical Society of Japan}\ }\textbf {\bibinfo
  {volume} {88}},\ \bibinfo {pages} {063706} (\bibinfo {year}
  {2019})}\BibitemShut {NoStop}%
\bibitem [{\citenamefont {Hafner}\ \emph {et~al.}(2019)\citenamefont {Hafner},
  \citenamefont {Rai}, \citenamefont {Banda}, \citenamefont {Kliemt},
  \citenamefont {Krellner}, \citenamefont {Sichelschmidt}, \citenamefont
  {Morosan}, \citenamefont {Geibel},\ and\ \citenamefont {Brando}}]{Brando}%
  \BibitemOpen
  \bibfield  {author} {\bibinfo {author} {\bibfnamefont {D.}~\bibnamefont
  {Hafner}}, \bibinfo {author} {\bibfnamefont {B.~K.}\ \bibnamefont {Rai}},
  \bibinfo {author} {\bibfnamefont {J.}~\bibnamefont {Banda}}, \bibinfo
  {author} {\bibfnamefont {K.}~\bibnamefont {Kliemt}}, \bibinfo {author}
  {\bibfnamefont {C.}~\bibnamefont {Krellner}}, \bibinfo {author}
  {\bibfnamefont {J.}~\bibnamefont {Sichelschmidt}}, \bibinfo {author}
  {\bibfnamefont {E.}~\bibnamefont {Morosan}}, \bibinfo {author} {\bibfnamefont
  {C.}~\bibnamefont {Geibel}},\ and\ \bibinfo {author} {\bibfnamefont
  {M.}~\bibnamefont {Brando}},\ }\bibfield  {title} {\bibinfo {title}
  {Kondo-lattice ferromagnets and their peculiar order along the magnetically
  hard axis determined by the crystalline electric field},\ }\href
  {https://doi.org/10.1103/PhysRevB.99.201109} {\bibfield  {journal} {\bibinfo
  {journal} {Phys. Rev. B}\ }\textbf {\bibinfo {volume} {99}},\ \bibinfo
  {pages} {201109} (\bibinfo {year} {2019})}\BibitemShut {NoStop}%
\bibitem [{\citenamefont {Nakatsuji}\ \emph {et~al.}(2004)\citenamefont
  {Nakatsuji}, \citenamefont {Pines},\ and\ \citenamefont
  {Fisk}}]{Nakatsuji2004Jan}%
  \BibitemOpen
  \bibfield  {author} {\bibinfo {author} {\bibfnamefont {S.}~\bibnamefont
  {Nakatsuji}}, \bibinfo {author} {\bibfnamefont {D.}~\bibnamefont {Pines}},\
  and\ \bibinfo {author} {\bibfnamefont {Z.}~\bibnamefont {Fisk}},\ }\bibfield
  {title} {\bibinfo {title} {{Two Fluid Description of the Kondo Lattice}},\
  }\href {https://doi.org/10.1103/PhysRevLett.92.016401} {\bibfield  {journal}
  {\bibinfo  {journal} {Phys. Rev. Lett.}\ }\textbf {\bibinfo {volume} {92}},\
  \bibinfo {pages} {016401} (\bibinfo {year} {2004})}\BibitemShut {NoStop}%
\bibitem [{\citenamefont {Yang}\ and\ \citenamefont
  {Pines}(2008)}]{Yang2008Mar}%
  \BibitemOpen
  \bibfield  {author} {\bibinfo {author} {\bibfnamefont {Y.-f.}\ \bibnamefont
  {Yang}}\ and\ \bibinfo {author} {\bibfnamefont {D.}~\bibnamefont {Pines}},\
  }\bibfield  {title} {\bibinfo {title} {{Universal Behavior in Heavy-Electron
  Materials}},\ }\href {https://doi.org/10.1103/PhysRevLett.100.096404}
  {\bibfield  {journal} {\bibinfo  {journal} {Phys. Rev. Lett.}\ }\textbf
  {\bibinfo {volume} {100}},\ \bibinfo {pages} {096404} (\bibinfo {year}
  {2008})}\BibitemShut {NoStop}%
\bibitem [{\citenamefont {Curro}\ \emph {et~al.}(2004)\citenamefont {Curro},
  \citenamefont {Young}, \citenamefont {Schmalian},\ and\ \citenamefont
  {Pines}}]{Schmallian_knight}%
  \BibitemOpen
  \bibfield  {author} {\bibinfo {author} {\bibfnamefont {N.~J.}\ \bibnamefont
  {Curro}}, \bibinfo {author} {\bibfnamefont {B.-L.}\ \bibnamefont {Young}},
  \bibinfo {author} {\bibfnamefont {J.}~\bibnamefont {Schmalian}},\ and\
  \bibinfo {author} {\bibfnamefont {D.}~\bibnamefont {Pines}},\ }\bibfield
  {title} {\bibinfo {title} {Scaling in the emergent behavior of heavy-electron
  materials},\ }\href {https://doi.org/10.1103/PhysRevB.70.235117} {\bibfield
  {journal} {\bibinfo  {journal} {Phys. Rev. B}\ }\textbf {\bibinfo {volume}
  {70}},\ \bibinfo {pages} {235117} (\bibinfo {year} {2004})}\BibitemShut
  {NoStop}%
\bibitem [{\citenamefont {Kwasigroch}\ \emph {et~al.}(2022)\citenamefont
  {Kwasigroch}, \citenamefont {Hu}, \citenamefont
  {Kr{\ifmmode\ddot{u}\else\"{u}\fi}ger},\ and\ \citenamefont
  {Green}}]{Kwasigroch2022Jun}%
  \BibitemOpen
  \bibfield  {author} {\bibinfo {author} {\bibfnamefont {M.~P.}\ \bibnamefont
  {Kwasigroch}}, \bibinfo {author} {\bibfnamefont {H.}~\bibnamefont {Hu}},
  \bibinfo {author} {\bibfnamefont {F.}~\bibnamefont
  {Kr{\ifmmode\ddot{u}\else\"{u}\fi}ger}},\ and\ \bibinfo {author}
  {\bibfnamefont {A.~G.}\ \bibnamefont {Green}},\ }\bibfield  {title} {\bibinfo
  {title} {{Magnetic hard-direction ordering in anisotropic Kondo systems}},\
  }\href {https://doi.org/10.1103/PhysRevB.105.224418} {\bibfield  {journal}
  {\bibinfo  {journal} {Phys. Rev. B}\ }\textbf {\bibinfo {volume} {105}},\
  \bibinfo {pages} {224418} (\bibinfo {year} {2022})}\BibitemShut {NoStop}%
\bibitem [{\citenamefont {Andrade}\ \emph {et~al.}(2014)\citenamefont
  {Andrade}, \citenamefont {Brando}, \citenamefont {Geibel},\ and\
  \citenamefont {Vojta}}]{Vojta&Brando}%
  \BibitemOpen
  \bibfield  {author} {\bibinfo {author} {\bibfnamefont {E.~C.}\ \bibnamefont
  {Andrade}}, \bibinfo {author} {\bibfnamefont {M.}~\bibnamefont {Brando}},
  \bibinfo {author} {\bibfnamefont {C.}~\bibnamefont {Geibel}},\ and\ \bibinfo
  {author} {\bibfnamefont {M.}~\bibnamefont {Vojta}},\ }\bibfield  {title}
  {\bibinfo {title} {Competing orders, competing anisotropies, and
  multicriticality: The case of co-doped
  ${\mathrm{ybrh}}_{2}{\mathrm{si}}_{2}$},\ }\href
  {https://doi.org/10.1103/PhysRevB.90.075138} {\bibfield  {journal} {\bibinfo
  {journal} {Phys. Rev. B}\ }\textbf {\bibinfo {volume} {90}},\ \bibinfo
  {pages} {075138} (\bibinfo {year} {2014})}\BibitemShut {NoStop}%
\bibitem [{\citenamefont {Scott}\ and\ \citenamefont
  {Kwasigroch}(2025)}]{Scott2025Dec}%
  \BibitemOpen
  \bibfield  {author} {\bibinfo {author} {\bibfnamefont {E.}~\bibnamefont
  {Scott}}\ and\ \bibinfo {author} {\bibfnamefont {M.}~\bibnamefont
  {Kwasigroch}},\ }\bibfield  {title} {\bibinfo {title} {{Destabilization of
  local magnetic anisotropy in heavy-fermion compounds}},\ }\href
  {https://doi.org/10.1103/7zsl-4497} {\bibfield  {journal} {\bibinfo
  {journal} {Phys. Rev. Res.}\ }\textbf {\bibinfo {volume} {7}},\ \bibinfo
  {pages} {043225} (\bibinfo {year} {2025})}\BibitemShut {NoStop}%
\bibitem [{\citenamefont {Jin}\ \emph {et~al.}(2025)\citenamefont {Jin},
  \citenamefont {Ullah}, \citenamefont {Klavins},\ and\ \citenamefont
  {Taufour}}]{Jin2025}%
  \BibitemOpen
  \bibfield  {author} {\bibinfo {author} {\bibfnamefont {H.}~\bibnamefont
  {Jin}}, \bibinfo {author} {\bibfnamefont {R.~R.}\ \bibnamefont {Ullah}},
  \bibinfo {author} {\bibfnamefont {P.}~\bibnamefont {Klavins}},\ and\ \bibinfo
  {author} {\bibfnamefont {V.}~\bibnamefont {Taufour}},\ }\bibfield  {title}
  {\bibinfo {title} {Importance of anisotropic interactions for hard-axis or
  hard-plane ordering of ce-based ferromagnets},\ }\href
  {https://doi.org/10.1103/99ww-msjh} {\bibfield  {journal} {\bibinfo
  {journal} {Phys. Rev. B}\ }\textbf {\bibinfo {volume} {112}},\ \bibinfo
  {pages} {024437} (\bibinfo {year} {2025})}\BibitemShut {NoStop}%
\bibitem [{\citenamefont {Bulla}\ \emph {et~al.}(2008)\citenamefont {Bulla},
  \citenamefont {Costi},\ and\ \citenamefont {Pruschke}}]{NRG_RMP}%
  \BibitemOpen
  \bibfield  {author} {\bibinfo {author} {\bibfnamefont {R.}~\bibnamefont
  {Bulla}}, \bibinfo {author} {\bibfnamefont {T.~A.}\ \bibnamefont {Costi}},\
  and\ \bibinfo {author} {\bibfnamefont {T.}~\bibnamefont {Pruschke}},\
  }\bibfield  {title} {\bibinfo {title} {{Numerical renormalization group
  method for quantum impurity systems}},\ }\href
  {https://doi.org/10.1103/RevModPhys.80.395} {\bibfield  {journal} {\bibinfo
  {journal} {Rev. Mod. Phys.}\ }\textbf {\bibinfo {volume} {80}},\ \bibinfo
  {pages} {395} (\bibinfo {year} {2008})}\BibitemShut {NoStop}%
\bibitem [{\citenamefont {Legeza}\ \emph {et~al.}(2008)\citenamefont {Legeza},
  \citenamefont {Moca}, \citenamefont {Toth}, \citenamefont {Weymann},\ and\
  \citenamefont {Zarand}}]{fnrg}%
  \BibitemOpen
  \bibfield  {author} {\bibinfo {author} {\bibfnamefont {O.}~\bibnamefont
  {Legeza}}, \bibinfo {author} {\bibfnamefont {C.~P.}\ \bibnamefont {Moca}},
  \bibinfo {author} {\bibfnamefont {A.~I.}\ \bibnamefont {Toth}}, \bibinfo
  {author} {\bibfnamefont {I.}~\bibnamefont {Weymann}},\ and\ \bibinfo {author}
  {\bibfnamefont {G.}~\bibnamefont {Zarand}},\ }\bibfield  {title} {\bibinfo
  {title} {{Manual for the Flexible DM-NRG code}},\ }\href
  {https://arxiv.org/abs/0809.3143} {\bibfield  {journal} {\bibinfo  {journal}
  {arXiv:0809.3143}\ } (\bibinfo {year} {2008})},\ \bibinfo {note} {the code is
  available at \url{http://www.phy.bme.hu/~dmnrg/}}\BibitemShut {NoStop}%
\bibitem [{\citenamefont {Rajan}(1983)}]{Rajan1983Jul}%
  \BibitemOpen
  \bibfield  {author} {\bibinfo {author} {\bibfnamefont {V.~T.}\ \bibnamefont
  {Rajan}},\ }\bibfield  {title} {\bibinfo {title} {{Magnetic Susceptibility
  and Specific Heat of the Coqblin-Schrieffer Model}},\ }\href
  {https://doi.org/10.1103/PhysRevLett.51.308} {\bibfield  {journal} {\bibinfo
  {journal} {Phys. Rev. Lett.}\ }\textbf {\bibinfo {volume} {51}},\ \bibinfo
  {pages} {308} (\bibinfo {year} {1983})}\BibitemShut {NoStop}%
\bibitem [{\citenamefont {W\'{o}jcik}\ \emph {et~al.}()\citenamefont
  {W\'{o}jcik}, \citenamefont {Kroha},\ and\ \citenamefont {Wahl}}]{KWJKPW}%
  \BibitemOpen
  \bibfield  {author} {\bibinfo {author} {\bibfnamefont {K.~P.}\ \bibnamefont
  {W\'{o}jcik}}, \bibinfo {author} {\bibfnamefont {J.}~\bibnamefont {Kroha}},\
  and\ \bibinfo {author} {\bibfnamefont {P.}~\bibnamefont {Wahl}},\ }\href@noop
  {} {\ }\bibinfo {note} {In preparation}\BibitemShut {NoStop}%
\bibitem [{\citenamefont {Georges}\ \emph {et~al.}(1996)\citenamefont
  {Georges}, \citenamefont {Kotliar}, \citenamefont {Krauth},\ and\
  \citenamefont {Rozenberg}}]{DMFT_RMP}%
  \BibitemOpen
  \bibfield  {author} {\bibinfo {author} {\bibfnamefont {A.}~\bibnamefont
  {Georges}}, \bibinfo {author} {\bibfnamefont {G.}~\bibnamefont {Kotliar}},
  \bibinfo {author} {\bibfnamefont {W.}~\bibnamefont {Krauth}},\ and\ \bibinfo
  {author} {\bibfnamefont {M.~J.}\ \bibnamefont {Rozenberg}},\ }\bibfield
  {title} {\bibinfo {title} {{Dynamical mean-field theory of strongly
  correlated fermion systems and the limit of infinite dimensions}},\ }\href
  {https://doi.org/10.1103/RevModPhys.68.13} {\bibfield  {journal} {\bibinfo
  {journal} {Rev. Mod. Phys.}\ }\textbf {\bibinfo {volume} {68}},\ \bibinfo
  {pages} {13} (\bibinfo {year} {1996})}\BibitemShut {NoStop}%
\end{thebibliography}

%

\end{document}